\documentclass[twocolumn,showpacs,
amsmath,amssymb]{revtex4}
\usepackage{amssymb}
\usepackage[dvips]{graphicx}

\begin{document}

\title{Electron relaxation in metals: Theory and exact analytical
    solutions}
\author{V. V. Kabanov$^{1,2}$ and A. S. Alexandrov$^{2,1}$ }

\affiliation{
$^1$Josef Stefan Institute 1001, Ljubljana, Slovenia \\
$^2$Department of Physics, Loughborough University,
Loughborough LE11 3TU, United Kingdom}

\begin{abstract}
The non-equilibrium dynamics of electrons is of a great experimental and theoretical value providing important microscopic parameters of the  Coulomb and  electron-phonon interactions  in metals and other cold plasmas. Because of the mathematical complexity of  collision integrals theories of  electron relaxation  often rely  on the assumption that electrons are in a "quasi-equilibrium"  (QE) with a time-dependent temperature, or on the numerical integration
 of the time-dependent Boltzmann equation.
We transform the integral Boltzmann equation to a partial
differential
  Schr\"odinger-like  equation with  imaginary time
in a one-dimensional "coordinate"  space reciprocal to energy which
allows for exact analytical solutions in both cases of
electron-electron and electron-phonon relaxation. The exact
relaxation rates are compared with the QE relaxation rates at high
and low temperatures.
\end{abstract}


\maketitle

\section{Introduction}
In recent years  investigations of  photo-response functions in advanced materials  have gone through a vigorous revival. In particular, laser "pump-probe" techniques, where a second probe pulse is delayed in time with respect to the pump pulse,
provide unique information on the strength of electron-electron (e-e) and electron-phonon (e-ph) interactions in metals and doped insulators
if an adequate theory is in place.

At present  detailed experimental data on relaxation processes is collected
for metals
\cite{brorson,schoenlein,elsayed,groeneveld1,brorson1}
and  high-temperature superconductors \cite{eesley,han,chekalin,albrecht,stevens,demsar,kabanov}.
The pump-probe experiments  are routinely analyzed in the framework of the so-called two
temperature model (TTM) \cite{kag,allen}. The  model  is based on the assumption that electrons
and phonons are in a thermal quasi-equilibrium (QE)  with two different time-dependent
temperatures $T_{e}(t)$ and $T_{l}(t)$, respectively. The  comprehensive
analysis of  experimental data collected at  room temperature  \cite{brorson}  allowed for a determination of the electron-phonon (e-ph)
coupling constant $\lambda$ of many metals and low-temperature superconductors in the framework of TTM.

Similar experiments and their analysis were performed on high-temperature
superconductors.  The femtosecond time-resolved measurements  on the high-T$_c$ superconductors Tl$_2$Ba$_2$Ca$_2$Cu$_3$O$_{10}$ \cite{eesley}  and YBa$_{2}$Cu$_3$O$_{7-\delta}$ \cite{han,chekalin}  found a relaxation process below T$_c$, which is distinct from the equilibration of hot carriers in the normal state.
A relatively strong e-ph coupling, $\lambda\simeq 0.9$
 \cite{chekalin}, a rapid decrease of the photo-response decay rate  with decreasing temperature \cite{segre} were found  in  YBa$_{2}$Cu$_3$O$_{6.5}$,  and
 the phonon bottleneck  \cite{rt,kabanov,kabanov1,demsar1} or a biparticle
recombination \cite{segre,kaindl} were observed below T$_c$.
 More
recently a time-resolved photoemission spectroscopy
\cite{perfetti} and the standard pump-probe optical measurements \cite{zhu} have been performed on
Bi$_{2}$Sr$_{2}$CaCu$_{2}$O$_{8+\delta}$. Their TTM analysis  has led to a rather weak e-ph coupling, $\lambda < 0.25$.

The pump-probe techniques have a potential to resolve a
controversial issue on weather  the e-ph interaction  is crucial
\cite{alemot} or weak and inessential \cite{and} for the mechanism
of high-temperature superconductivity. The pioneering work by
Kaganov, Lifshitz and Tanatarov \cite{kag} and subsequent TTM
studies are based on the assumption that  electrons are in the
thermal QE-state because the e-e relaxation time is supposed to be
much shorter than the e-ph relaxation time. This assumption is of
course incorrect on a femtosecond scale comparable with the e-e
scattering time of highly excited electrons, but the expectation has
been  that deviation from QE may not in fact have much influence on
the electron energy relaxation on a larger time scale \cite{allen}
(for  discussions of TTM with respect to some experiments see for
example \cite{demsar3,demsar4}).

However later on it has been realized that nonthermal effects are
essential even on a picosecond scale, comparable with the e-ph
relaxation time, when conditions of low laser excitation power and
relatively low temperature are chosen \cite{lag}. Under these
conditions, the e-e collision rate becomes strongly suppressed as a
result of the Pauli exclusion principle. Numerically integrating the
Boltzmann equation with  e-e and e-ph collision integrals
Groeneveld, Sprik, and Lagendijk  \cite{lag} have shown that the
electron gas cannot attain a thermal distribution by e-e collisions
on the time scale of the e-ph energy relaxation. A departure from QE
leads to an increase of the e-ph energy relaxation time with respect
to the QE expectation. As a consequence of this departure  one might
underestimate the e-ph coupling using TTM.

 While  numerical integrations of the Boltzmann equation can describe the time evolution of the
electron distribution function on any time scale, they require a
number of input parameters, which might be  unknown a priori.  Here
an analytical approach to this long-standing problem is developed.
We reduce  the integral Boltzmann equation to a differential
Schr\"odinger-like  equation using an auxiliary  space reciprocal to
energy and find exact analytical time-dependent distributions of
electrons in both cases of  electron-electron  and  electron-phonon
relaxation. We also derive  long-time relaxation rates of response
functions and compare them  with TTM.

\section{Electron-electron relaxation}\label{sec2}
Let us first consider a nonthermal relaxation of the electron distribution function $f_{\bf k}(t)$ caused by  electron-electron collisions, which is described by the following Boltzmann equation,
\begin{eqnarray}
\dot{f}_{\bf k}&=&{2\pi\over{\hbar}} \sum_{\bf p,q} V_c^2({\bf q})\delta(\xi_{\bf k}+\xi_{\bf
p}-\xi_{\bf k+q}-\xi_{\bf
p-q})[f_{\bf k+q}f_{\bf p-q} \cr
&\times& (1-f_{\bf
k})(1-f_{\bf p}) - f_{\bf k}f_{\bf p}(1-f_{\bf
k+q})(1-f_{\bf p-q})].
\end{eqnarray}
Here $\dot{f}_{\bf k}\equiv \partial f_{\bf k}(t)/\partial t$,
$V_c({\bf q})$ is the matrix element of the electron-electron
scattering (pseudo)potential, $\xi_{\bf k}$ is the electron energy
with respect to the equilibrium chemical potential. For transparency
we drop the time argument in the distribution function. If the
distribution function depends only on energy and time, $f_{\bf
k}=f_{\xi}$,
 one can
average this equation over the  angles of ${\bf k}$ as
\begin{equation}
\dot{f}_\xi\equiv N^{-1}(\xi)\sum_{\bf k} \delta(\xi_{\bf
k}-\xi)\dot{f}_{\bf k},\label{average}
\end{equation}
where $N(\xi)$ is the density of states  (DOS) per spin, with the following result
\begin{eqnarray}
&&\dot{f}_\xi=\int\int \int d\xi' d\epsilon d\epsilon'
K(\xi,\xi',\epsilon,\epsilon')\delta(\xi+\epsilon-\xi'-\epsilon')\times \cr
&& \left[f_{\xi'}f_{\epsilon'}(1-f_\xi)(1-f_\epsilon) - f_\xi f_\epsilon (1-f_{\xi'})(1-f_{\epsilon'})\right],
\label{ecollision2}
\end{eqnarray}
where
\begin{eqnarray}
K(\xi,\xi', \epsilon, \epsilon')&=&{2\pi\over{\hbar N(\xi)}}\sum_{\bf k,p,q}  V_c^2({\bf q})
\delta(\xi_{\bf k}-\xi)\times \cr
&& \delta(\xi_{\bf p}-\epsilon) \delta(\xi_{\bf
k+q}-\xi')\delta(\xi_{\bf
p-q}-\epsilon').
\end{eqnarray}

We restrict our theory to relaxations  involving non-equilibrium
electron-hole excitations with energies much less than the
equilibrium Fermi energy, $E_F$. Since the kernel $K(\xi,\xi',
\epsilon, \epsilon')$ has variation on a scale of  the Fermi energy,
one can approximate it by a constant, $K(\xi,\xi', \epsilon,
\epsilon')\approx K$. This constant is related to the Coulomb
pseudo-potential $\mu_c\equiv V_cN(0)$, important in the  theory of
superconductivity,  $K\approx \pi \mu_c^2/2\hbar E_F$. Assuming a
low laser excitation power we linearize Eq.(\ref{ecollision2}) by
introducing a small non-equilibrium correction, $\phi(\xi,t)\ll 1$,
to the  equilibrium distribution, $n_{\xi}$,
\begin{equation}
f_\xi=n_{\xi}+ \phi(\xi,t)\label{lin}
\end{equation}
where $n_{\xi}=(e^{\xi/k_BT}+1)^{-1}$. Keeping terms linear in $\phi(\xi,t)$ and  measuring  energies  in units of $k_BT$, which is the only relevant energy scale of the problem, one obtains
\begin{eqnarray}
\dot{\phi} (\xi,t)&=&K (k_BT)^2\int\int \int d\xi' d\epsilon d\epsilon'
\delta(\xi+\epsilon-\xi'-\epsilon')\times \cr
&& [\phi(\xi',t)(n_{-\xi}n_{-\epsilon}n_{\epsilon'}+n_{\xi}n_{\epsilon}n_{-\epsilon'})\cr
&-&
\phi(\xi,t)(n_{\xi'}n_{-\epsilon}n_{\epsilon'}+n_{-\xi'}n_{\epsilon}n_{-\epsilon'})].
\label{ecollision2l}
\end{eqnarray}

Performing simple integrations in  linearized Eq.(\ref{ecollision2l}) yields
\begin{eqnarray}
&&\dot{\phi}(\xi,t)=-{\phi(\xi,t)\over{\tau_e(\xi)}}+
{K(k_BT)^2\over{\cosh(\xi/2)}}\int_{-\infty}^{\infty} d\xi'\phi(\xi',t)\times \cr
&& \cosh(\xi'/2)\left[{\xi-\xi' \over{\sinh({\xi-\xi'\over{2}})}}-{\xi+\xi' \over{2\sinh({\xi+\xi'\over{2}})}}\right] ,
\label{ecollision3}
\end{eqnarray}
where
\begin{equation}
\tau_e(\xi)= {2\over{(\pi^2+\xi^2)K(k_BT)^2}} \label{taue}
 \end{equation}
 is  the familiar lifetime of  electron-hole excitations in the Fermi liquid. Here we have used the integral $\int_{0}^{\infty} dz \ln (z)/[(z-a)(z+b)]= [\pi^2-\ln^2(a)+\ln^2(b)]/2(a+b)$ with  $a,b >0$.

The second term on the right-hand side of Eq.(\ref{ecollision3})
describes a source of quasi-particles due to inelastic
electron-electron collisions. Collisions in cold degenerate plasmas
differ essentially from quasi-elastic collisions in classical (hot)
plasmas. In the latter the energy transfer is small compared with
the electron energy due to a  long-range character of the Coulomb
potential, so that one can approximate the  Boltzmann collision
integral by the differential Landau-Fokker-Plank (LFP) equation
(see, for example Ref.\cite{karas}).  As one can see  from
Eq.(\ref{ecollision3}) it is not the case in metals. The collision
energy transfer in metals is about the same as the excitation energy
itself, which makes the differential LFP approximation unacceptable
here.

Remarkably the electron-electron collision integral  acquires a differential form in a reciprocal auxiliary-time space introduced via the Fourier transform of Eq.(\ref{ecollision3}), rather than in the energy space as in the LFP case. Let us consider non-equilibrium states conserving the electron-hole symmetry, so that the non-equilibrium part of the distribution is an odd function of energy, $\phi(-\xi,t)=- \phi(\xi,t)$.  If one determines a  function,
\begin{equation}
\chi(\xi,t)\equiv \phi(\xi,t) \cosh(\xi/2),\label{def}
\end{equation}
then  the Boltzmann equation is simplified as
\begin{equation}
{\dot{\chi}(\xi,t)\over{K(k_BT)^2}}=-{{\pi^2+\xi^2}\over{2}}\chi(\xi,t)+
{3\over{2}}\int_{-\infty}^{\infty} d\xi'\chi(\xi',t)
{\xi-\xi' \over{\sinh({\xi-\xi'\over{2}})}} .
\label{ecollision4}
\end{equation}
We shall see below that a "bound state"  of the effective
"Schr\"odinger" equation for  the Fourier transform of $\chi(\xi,t)$
corresponds to the stationary quasi-equilibrium distribution.
Taking the Fourier transform of Eq.(\ref{ecollision4}), we arrive at
an exact differential counterpart of the Boltzmann equation,
\begin{equation}
\tau_{e}\dot{\psi}(x,t)=\left [{\partial^2\over{\partial x^2}} +{6\over{\cosh^2(x)}}-1 \right]\psi(x,t),
\label{soliton}
\end{equation}
where
\begin{equation}
\tau_e=  2/\pi^2K(k_BT)^2 \label{taue2}
\end{equation}
and
\begin{equation}
\psi(x,t)=\int_{-\infty}^{\infty} d \xi \chi(\xi,t)e^{ix\xi/\pi}.\label{fourier}
\end{equation}
Here another integral $\int_{-\infty}^{\infty}dz z \exp(ix z)/\sinh(z/2)= 2\pi^2/\cosh^2(\pi x)$ has been used.

The  solution of Eq.(\ref{soliton}) is found as a superposition of  normalized eigenstates, $\psi_k(x)$, of a textbook Hamiltonian \cite{landau},
\begin{equation}
\psi(x,t)= \sum_{k}  c_k \psi_k(x)e^{(k^2-1)t/\tau_{e}}, \label{exact}
\end{equation}
where  coefficients $c_k$ are determined by the initial non-equilibrium distribution function $\phi(\xi,0)$ at $t=0$,
\begin{equation}
c_k= \int_{-\infty}^{\infty}dx\int_{-\infty}^{\infty}d\xi \psi^{*}_k(x)\cosh(\xi/2)\phi(\xi,0)e^{ix\xi/\pi}.
 \label{coefficient}
\end{equation}
The eigenstates, $\psi_k(x)$, and the eigenvalues,  $E=-k^2$, are found from the Schr\"odinger equation
\begin{equation}
\left [{\partial^2\over{\partial x^2}} +{m(m+1)\over{\cosh^2(x)}} \right]\psi_k(x)=k^2\psi_k(x),
\label{soliton2}
\end{equation}
with $m=2$. This equation has a finite number of discrete bound states  with real $k$'s \cite{landau} and  continuum extended states  with imaginary $k$'s, $k=ip$ ($p$ is  real ). If $m$ is an integer there are  $m$ bound states with $k=1,2,..., m$ and both bound and unbound eigenstates can be expressed in terms of elementary functions \cite{ves},
\begin{equation}
\psi_k(x)=A_k \hat{D}_m \hat{D}_{m-1}...\hat{D}_1 e^{kx},
\end{equation}
where $A_k$ is the normalizing amplitude and $\hat{D}_m=d/dx-m \tanh(x)$. In our case ($m=2$) there are two bound states, the even ground state with $k=2$ ($E=-4$) and the odd excited state with $k=1$($E=-1$). For relaxations conserving the electron-hole symmetry the ground state contribution to the superposition, Eq.(\ref{exact}), is integrated to zero because the initial non-equilibrium distribution is odd. On the contrary, the  excited odd state with $\psi_1(x) \propto \sinh(x)/\cosh^2(x)$ is the only state, which survives in Eq.(\ref{exact}) at $t \rightarrow \infty$, so that
\begin{equation}
\psi(x,\infty)\propto {\sinh(x)\over{\cosh^2(x)}},
\end{equation}
and (using Eq.(\ref{fourier}) and Eq.(\ref{def}))
\begin{equation}
\phi(x,\infty)\propto {\xi\over{\cosh^2(\xi/2)}},\label{correction}
\end{equation}
which is precisely the result of the QE approximation. Indeed expanding the QE distribution function, $f_{QE}=[\exp((E-E_F)/T_e)+1]^{-1}$ in powers of $T_e-T$, we obtain
\begin{equation}
f_{QE}=n_{\xi}+\left({T_e-T\over{4T}}\right){\xi\over{\cosh^2(\xi/2)}}\label{eq}
\end{equation}
with the same non-equilibrium correction, $n_{QE} \propto \xi/\cosh^2(\xi/2)$, as in Eq.(\ref{correction}).

\begin{figure}
\includegraphics[width = 87mm, angle=-0]{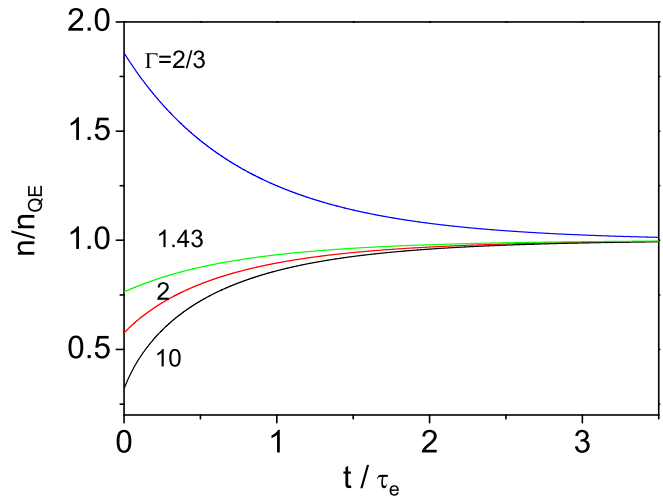}
\caption{Relaxation of the total number of non-equilibrium excitations $n$ normalized by the quasi-equilibrium number, $n_{QE}$, for different width $\Gamma$ of the initial non-equilibrium distribution.}
\end{figure}

The  exact solution, Eq.(\ref{exact}), allows us to trace the
relaxation at any energy and at any time scale. In particular  Fig.1
represents the time evolution of the  total number of
non-equilibrium excitations (electrons plus holes),
$n=2\int_{0}^{\infty}d\xi \phi(\xi,t)$, for  initial non-equilibrium
distributions of the shape, $\phi(\xi,0)=
(\pi^{1/2}E_0/2\Gamma^3)\xi \exp[-(\xi/\Gamma)^2]$, where the
distribution width  $\Gamma$ is varied, but the  total energy,
$E_0$, is  unchanged. For wide sources with large $\Gamma > 1 $ the
number of excitations increases with time conserving the total
energy in the process of their cooling. On the other hand, when most
excitations at $t=0$ are created  with the  energy less than $k_BT$
(i.e. $\Gamma < 1$), their number decreases with time since their
individual energies increase due to collisions with equilibrium
electrons. One can trace the relaxation for any  initial
distribution including the case when   a single photon initially
creates a single electron hole pair, which cascades eventually into
a large number  of pairs until they get lost in the background
thermal distribution.

The asymptotic behavior of response functions can be readily
obtained from  Eq.(\ref{exact}) by taking into account that only the excited bound state and the  extended states with small $p \approx (t/\tau_e)^{-1/2} \ll 1$ contribute  to the sum in Eq.(\ref{exact}), when $t/\tau_e \gg (1, 1/\Gamma^2)$. Substituting the  extended eigenfunctions, $\psi_p(x) \propto [3\tanh^2(x)-3ip \tanh(x)-p^2-1]\exp(ipx)$ into Eq.(\ref{exact}) with $c_p \propto p$ at small $p$  and integrating over $p$ yield
 \begin{equation}
 \psi(x,t)-\psi(x,\infty) \propto {x\over{t^{3/2}}} \exp \left(-{t\over\tau_e}-{x^2 \tau_e\over{4t}}\right).\label{asymp}
 \end{equation}
 in the saddle-point approximation. Performing the Fourier transform of Eq.(\ref{asymp}) with respect to $x$ we find
 \begin{equation}
 \phi(\xi,t)-\phi(\xi,\infty) \propto \xi e^{-t/\tau_e(\xi)}.\label{asymp2}
 \end{equation}
 The same  result is obtained by using a $\tau$-approximation for the Boltzmann equation (\ref{ecollision3}),
  \begin{equation}
\dot{\phi}(\xi,t)=-{\phi(\xi,t)-\phi(\xi, \infty)\over{\tau_e(\xi)}},
\label{tau}
\end{equation}
which has  the following solution
\begin{equation}
\phi(\xi,t)=\phi(\xi, \infty)+[\phi(\xi, 0)-\phi(\xi, \infty)]e^{-t/\tau_e(\xi)}. \label{tau2}
\end{equation}
Hence one can use the $\tau$-approximation, Eq.(\ref{tau2}), on the time scale much longer than the characteristic collision time, $\tau_e$. However  this approximation is inaccurate on a shorter time scale because in contrast with the exact solution, Eq.(\ref{exact}), it does not conserve the total energy.

 Integrating Eq.(\ref{asymp2}) yields a universal time-asymptotic of the total number of electron-hole excitations,
 \begin{equation}
n(t)-n_{QE} \propto {e^{-t/\tau_e}\over{t}}, \label{asymp3}
\end{equation}
 as also seen from Fig.1.

 Importantly the characteristic e-e relaxation time is quite long  due to the Pauli exclusion principle. Using  realistic $\mu_c=1$ and $E_{F}=10$ eV we estimate $\tau_e\approx 1.2$ ps   at the room temperature $T=300$ K (see also Ref.\cite{lag}), which increases further as $1/T^2$ with cooling.

\section{Electron-phonon relaxation}\label{sec3}
Now let us consider the electron-phonon relaxation   described by the e-ph collision integral,
\begin{eqnarray}
\dot{f}_{\bf k}&=&{2\pi\over{\hbar}}\sum_{\bf q} M^2({\bf q})\times \cr
&& \{[f_{\bf k-q}(1-f_{\bf k})N_{\bf q}-  f_{\bf k}(1-f_{\bf k-q})(N_{\bf q}+1)]\cr
&\times& \delta(\xi_{\bf k}-\xi_{\bf k-q}-\hbar \omega_{\bf q})+  \cr
&& [f_{\bf k+q}(1-f_{\bf k})(N_{\bf q}+1)-  f_{\bf k}(1-f_{\bf k+q})N_{\bf q}]\cr
 &\times& \delta(\xi_{\bf k}-\xi_{\bf k+q}+\hbar \omega_{\bf q})\}, \label{collision}
\end{eqnarray}
where $M({\bf q})$ is the matrix element of the deformation potential and $N_{\bf q}$ is the  distribution function of phonons with the frequency $\omega_{\bf q}$.

As in the former case of the e-e collisions we
average this equation over the momentum angles using Eq.(\ref{average}) and conventional units:
\begin{eqnarray}
\dot{f}_\xi&=&2\pi \int d\omega \int d\xi' Q(\omega,\xi,\xi')\times \cr
&& \{\delta(\xi-\xi'-\hbar\omega)[(f_{\xi'}-f_\xi)N_{\omega}-f_\xi (1-f_{\xi'})]\cr
&+& \delta(\xi-\xi'+\hbar\omega)[(f_{\xi'}-f_{\xi})N_{\omega}+f_{\xi'}(1-f_{\xi})]\}. \label{colaverage}
\end{eqnarray}
Here
\begin{equation}
Q(\omega,\xi,\xi')={1\over{\hbar N(\xi)}}\sum_{\bf k, q} M^2({\bf q})
\delta(\xi_{\bf k-q}-\xi')\delta(\xi_{\bf k}-\xi) \delta(\omega_{\bf
q}-\omega)
\end{equation}
is the e-ph spectral function \cite{allen}, which has $\omega$ variation on a scale of the maximum phonon frequency $\omega_D$ but $\xi$ and $\xi'$ variation only on a much larger energy scale of the order of $E_F$.

  Linearizing Eq.(\ref{colaverage}) with the help of Eq.(\ref{lin}) yields
\begin{eqnarray}
&&\dot{\phi}(\xi,t)=2\pi\int d\omega \int d\xi'
Q(\omega,\xi,\xi')\times \cr
&&\{[\phi(\xi',t)(N_{\omega}+n_{\xi})-\phi(\xi,t)(N_{\omega}+n_{-\xi'})] \delta(\xi-\xi'-\hbar\omega)+\cr
&&[\phi(\xi',t)(N_{\omega}+n_{-\xi})-\phi(\xi,t)(N_{\omega}+n_{\xi'})]\delta(\xi-\xi'+\hbar\omega)\}.\nonumber \\
\label{collision3}
\end{eqnarray}
Characteristic electron energies in
Eq.(\ref{collision3}) are much less than the Fermi energy, so that
\begin{equation}
Q(\omega,\xi,\xi')\approx Q(\omega,0,0)\equiv
\alpha^2 F(\omega)
\end{equation}
is the familiar Eliashberg function.  We also  assume that phonons are in the thermal equilibrium, $N_{\omega}= [\exp(\hbar \omega/k_BT)-1]^{-1}$, due to their fast thermalization  caused by anharmonic interactions (i.e.  phonon-phonon collisions) and/or due to a small  size of the sample and the pump-laser spot allowing for  a fast escape of non-equilibrium phonons. If this condition is not satisfied, one has to solve an equation for the non-equilibrium phonon distribution coupled with Eq.(\ref{collision3}), which is outside  the scope of this paper. Under these assumptions Eq.(\ref{collision3}) is transformed into a form similar to the e-e collision integral  Eq.(\ref{ecollision3}),
\begin{eqnarray}
\dot{\phi}(\xi,t)&=&-{\phi(\xi,t)\over{\tau_{ph}(\xi)}}+
{2\pi k_BT \over{\hbar \cosh(\xi/2)}}\int_{-\infty}^{\infty} d\xi' sign(\xi-\xi')\cr
&\times& \alpha^2 F\left({k_BT|\xi-\xi'|\over{\hbar}}\right) {\cosh(\xi'/2)\over{2\sinh({\xi-\xi'\over{2}})}}\phi(\xi',t),
\label{collision4}
\end{eqnarray}
where
\begin{eqnarray}
{1\over{\tau_{ph}(\xi)}}&=&{2\pi k_BT\over{\hbar}} \int_{0}^{\infty} d\omega \alpha^2 F\left({k_BT \omega\over{\hbar}}\right) \times \cr
&&\left[{1\over{\sinh({\omega\over{2}})\cosh({\omega\over{2}})}}+
{\sinh^2({\xi\over{2}})\tanh({\omega\over{2}}) \over{\cosh({\omega+\xi\over{2}})\cosh({\omega-\xi\over{2}})}}\right],\nonumber \\
\end{eqnarray}
and  energies are now measured in units of $k_BT$. The Eliashberg function is quite complicated in real metallic compounds because of their complex lattice structures. This complexity can be avoided in  a high-temperature regime, $k_BT \gg \hbar \omega_D$ and in an opposite low-temperature regime, $k_BT \ll \hbar \omega_D$.

\subsection{High-temperature electron-phonon relaxation} \label{sechigh}
As shown by Allen \cite{allen} the energy relaxation in TTM  has a particularly simple form in terms of the moments of $\alpha^2 F(\omega)$,
\begin{equation}
\lambda \langle \omega^n\rangle\equiv 2\int_{0}^{\infty}d\omega {\alpha^2 F(\omega)\omega^n\over {\omega}},
\end{equation}
where the  coupling constant $\lambda$, which determines the  critical temperature in the BCS superconductors, is
\begin{equation}
\lambda= 2\int_{0}^{\omega_D} d\omega {\alpha^2 F(\omega)\over{\omega}}.
\end{equation}
Here we derive a high-temperature LFP-type equation for the non-equilibrium part of the distribution function $\phi(\xi,t)$ to compare our exact approach with the TTM results \cite{kag,allen}.

At high temperatures the Eliashberg function is a narrow function on the temperature scale, so that one can apply a quasi-elastic approximation expanding the e-ph collision integral, Eq.(\ref{collision3}) or Eq.(\ref{collision4}), in powers of the  phonon energy, $\hbar \omega \ll k_BT$. The zero-order  elastic terms are canceled out because the distribution function depends  on energy only, while the next order terms yield the LFP-type differential equation,
\begin{equation}
 \gamma^{-1}\dot{\phi}(\xi,t) = {\partial\over{\partial \xi}} \left[\tanh(\xi/2) \phi(\xi)+{\partial \over{\partial \xi}}\phi(\xi)\right],
\label{high}
\end{equation}
where $\gamma=\pi \hbar \lambda \langle \omega^2\rangle/ k_BT$. Apart from a numerical coefficient of the order of $1$ the characteristic e-ph relaxation rate $\gamma/2$ is about the same as the TTM energy relaxation rate \cite{allen}, $\gamma_T=3\hbar \lambda \langle \omega^2\rangle/ \pi k_BT$, at high temperatures. Indeed multiplying Eq.(\ref{high}) by $\xi$ and integrating over all energies yield   the rate of excitation energy relaxation,
\begin{equation}
\dot{E}_e(t)=- \gamma \int_{-\infty}^{\infty} d\xi \tanh(\xi/2) \phi(\xi, t),
\label{rate}
\end{equation}
where $E_e(t)= \int_{-\infty}^{\infty} d\xi \xi \phi(\xi)$.  If we replace $\tanh(\xi/2)$ in this equation  by its argument assuming that $ \phi(\xi, t)$ has its characteristic energy width of the order of $1$, then $\dot{E}_e(t)\approx -(\gamma/2)E_e.$ Hence the excitation energy relaxes as  $E_e(t) \propto \exp(-\gamma t/2)$ almost independent on a particular  shape of the non-equilibrium distribution.

To verify the numerical coefficient, one can substitute  the TTM distribution $\phi_{TTM}(\xi,t)= [(T_e(t))-T)/4T]\xi/\cosh^2(\xi/2)$ (Eq.(\ref{eq})) into Eq.(\ref{rate}) to convert this into the temperature relaxation rate:
\begin{equation}
\dot{T}_e(t)=- {1\over{2}}\gamma (T_e-T) {\int_0^{\infty}dx x\tanh (x)/\cosh^2(x)\over{\int_0^{\infty}dx x^2/\cosh^2(x)}}.
\label{rate2}
\end{equation}
Eq.(\ref{rate2}) is precisely  the same as the TTM temperature rate \cite{allen}, $\dot{T}_e(t)=- \gamma_T (T_e-T)$, since the ratio of two integrals in Eq.(\ref{rate2}) is $6/\pi^2$.

According to Eq.(\ref{rate}) deviation of $\phi(\xi,t)$ from  quasi-equilibrium population does not have much influence on the energy relaxation. Hence TTM \cite{kag,allen} is the adequate approximation at high temperatures, which agrees well with experimental observations in conventional metals where Debye temperatures are rather low \cite{brorson}.

\subsection{Low-temperature electron-phonon relaxation in poor metals}
 Characteristic phonon frequencies are  exceptionally high in many advanced materials like copper oxides,  $\hbar \omega_D/k_B \gtrsim 400 \div 1000$ K, so that the low-temperature regime, $k_BT \ll \hbar \omega_D$ is of  great  importance.  Since all dimensionless energies in Eq.(\ref{collision4}) are of the order of unity one can apply  a low-frequency asymptotic of $\alpha^2 F(\omega)= \lambda n (\omega/\omega_D)^n/2$ in this regime. The exponent $n$ depends on impurities, disorder, and  sample dimensions: $n=2$ in clean bulk  crystals while $n=1$ in  disordered  metals  due to a  phonon damping  \cite{belitz,belitz2} and in  metallic films \cite{belevtsev}. Then  Eq.(\ref{collision4}) can be Fourier-transformed  into the Schr\"odinger equation using the Fourier transform Eq.(\ref{fourier}) of  $\chi(\xi,t)\equiv \phi(\xi,t) \cosh(\xi/2)$.

 In the poor-metal case $n=1$ the equation for $\chi(\xi,t)$ is almost the same as in the e-e case, Eq.(\ref{ecollision4}), apart from
a numerical coefficient in front of the integral term,
\begin{eqnarray}
&&{\dot{\chi}(\xi,t)\over{\pi\lambda(k_BT)^2/\hbar^2\omega_D}}=-{{\pi^2+\xi^2}\over{2}}\chi(\xi,t)+ \cr
&&{1\over{2}}\int_{-\infty}^{\infty} d\xi'\chi(\xi',t)
{\xi-\xi' \over{\sinh({\xi-\xi'\over{2}})}} ,
\label{collision5}
\end{eqnarray}
where the integrals
$\int_{0}^{\infty}d\omega
\omega/\sinh(\omega/2)\cosh(\omega/2)=\pi^2/2$ and $\int_{0}^{\infty}d \omega
\omega \tanh(\omega/2)/\cosh[(\omega+\xi)/2]\cosh[(\omega-\xi)/2]=\xi^2/2\sinh^2(\xi/2)$ have been used. The difference in the numerical coefficients in front of the integral terms originates in different statistics of scatterers, which are bosons in the e-ph case and fermions in the e-e case. At low temperatures the e-ph relaxation time has the same energy and temperature dependence as the e-e relaxation time Eq.(\ref{taue}),
\begin{equation}
\tau_{ph}(\xi)= {2\hbar^2\omega_D\over{(\pi^2+\xi^2)\pi\lambda(k_BT)^2}}. \label{tauph}
 \end{equation}
 We also notice that the temperature dependence of the e-ph relaxation rate at low temperatures, $1/\tau_{e-ph} \propto T^2$ is qualitatively different from its temperature dependence at high temperatures $\gamma \propto 1/T$.

The Fourier transform of Eq.(\ref{collision5}) yields the Schr\"odinger-like equation
\begin{equation}
\tau_{ph}\dot{\psi}(x,t)=\left [{\partial^2\over{\partial x^2}} +{2\over{\cosh^2(x)}}-1 \right]\psi(x,t),
\label{phsoliton}
\end{equation}
where
\begin{equation}
\tau_{ph}= {2\hbar^2 \omega_D\over{\pi^3\lambda(k_BT)^2}}.
\label{tauph2}
\end{equation}
Different from the e-e case [Eq.(\ref{soliton2}) with $m=2$] the steady-state Schr\"odinger equation
\begin{equation}
\left [{\partial^2\over{\partial x^2}} +{2\over{\cosh^2(x)}} \right]\psi_k(x)=k^2\psi_k(x),
\end{equation}
has only one  bound (ground) state, $k=1$,  and itinerant states with $k=ip$,
\begin{equation}
\psi_k(x)=A_k [k-\tanh(x)] e^{kx}
\end{equation}
in the e-ph case ($m=1$).
Only itinerant states contribute to the superposition Eq.(\ref{exact}) and determine the time relaxation of the distribution function because the contribution of the even ground state  is integrated  to zero and there is no excited odd state here. As the result the non-equilibrium part of the distribution function and the number of excitations relax  with characteristic time $\tau_{ph}$ to zero rather than to any quasi-equilibrium state  as shown in Fig.3 and Fig.4 by lower curves ($\alpha=2$). Their time asymptotic is found using the saddle-point approximation as in the case of the e-e collisions,
\begin{equation}
 \phi(\xi,t) \propto \xi e^{-t/\tau_{ph}(\xi)},\label{asymp3}
 \end{equation}
and
\begin{equation}
n(t) \propto {e^{-t/\tau_{ph}}\over{t}}. \label{asymp4}
\end{equation}
The time evolution of $n(t)$ is widely independent of the width $\Gamma$ of the initial distribution function at $t=0$ as one can see  comparing the lowest curves in Fig.3 and Fig.4.

\subsection{Low-temperature electron-phonon relaxation in clean metals}
The low-frequency Eliashberg function is quadratic as a function of  frequency, $\alpha^2 F(\omega)= \lambda (\omega/\omega_D)^2$ in clean crystalline metals, which makes an analytical expression  for the Fourier transform of the Boltzmann equation (\ref{collision4})  unavailable in terms of  elementary functions. However we can approximate all relevant integrals numerically as
\begin{equation}
\int_0^{\infty} d\omega {\omega^2 \over{\sinh(\omega/2)\cosh(\omega/2)}}\approx 8.414, \nonumber
\end{equation}
\begin{equation}
\int_{0}^{\infty} d\omega {\omega^2\sinh^2(\xi/2)\tanh(\omega/2) \over{\cosh({\omega+\xi\over{2}})\cosh({\omega-\xi\over{2}})}}\approx {3\over{2}}\xi^2+0.027 \xi^4, \nonumber
 \end{equation}
and
\begin{equation}
 \int_{0}^{\infty} d\omega {\omega^2\cos(\omega x/\pi) \over{\sinh(\omega/2)}}\approx {3\pi^2\over{2}}V(x). \nonumber
 \end{equation}
 where $V(x)$ is shown in Fig.2. Then the corresponding Schr\"odinger-type equation for the Fourier transform of $\chi(\xi,t)$ becomes
 \begin{equation}
\tau^{cl}_{ph}\dot{\psi}(x,t)=\left [{\partial^2\over{\partial x^2}}+0.178 {\partial^4\over{\partial x^4}} +V(x)-0.568 \right]\psi(x,t),
\label{phsoliton2}
\end{equation}
where now
\begin{equation}
\tau^{cl}_{ph}= {\hbar^3 \omega^2_D\over{3\pi^3\lambda(k_BT)^3}}.
\label{tauph3}
\end{equation}

\begin{figure}
\includegraphics[width = 87mm, angle=-0]{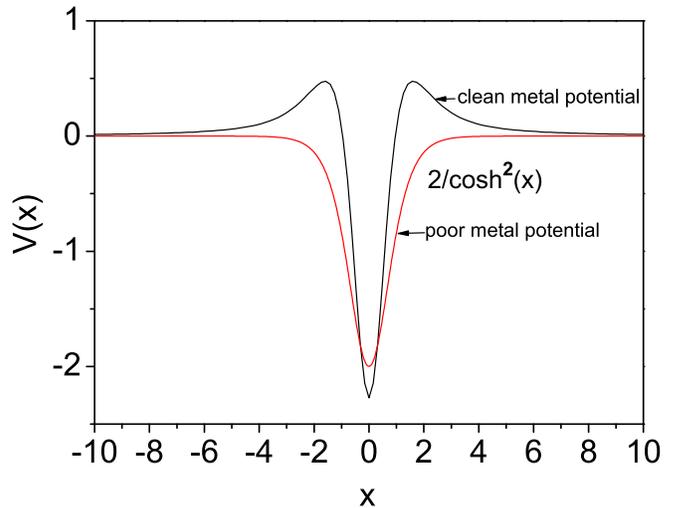}
\caption{The effective "potential" energy of the Schr\"odinger counterpart of the Boltzmann equation with the e-ph collision integral in clean metals compared with the potential in poor metals.}
\end{figure}

The effective "potential" energy differs only marginally from the
poor metal case, Fig.2. At large $t$ corresponding to large $x$ in
Eq.(\ref{phsoliton2}) the forth derivative of the low-energy
extended eigenstates is small. Hence   the asymptotic behavior of
response functions in clean metals is qualitatively about the same
as in poor metals,
\begin{equation}
n(t) \propto {\exp(-0.568 t/\tau^{cl}_{ph})\over{t}}, \label{asymp5}
\end{equation}
but the temperature dependence of the e-ph relaxation time is more
pronounced, $\tau^{cl}_{ph}\propto 1/T^3$. In principle, the
clean-metal "potential", Fig.2,  could have "resonances", states
that are in the continuum but take a long time to leak out resulting
in some quantitative differences with the poor-metal relaxation.

\section{Low-temperature electron-phonon relaxation combined with electron-electron relaxation}
Finally let us combine both collision integrals into one Boltzmann equation. Performing its Fourier transformation as described above in Sections (\ref{sec2}, \ref{sec3}) yields the following Schr\"odinger-like  equation in the poor-metal case:
\begin{equation}
\tau \dot{\psi}(x,t)=\left [{\partial^2\over{\partial x^2}} +{\alpha\over{\cosh^2(x)}}-1 \right]\psi(x,t),
\label{solitonf}
\end{equation}
where
\begin{equation}
\tau=  {\tau_e \tau_{ph}\over{\tau_e+\tau_{ph}}} \label{tauf}
\end{equation}
and
\begin{equation}
\alpha= {2\tau_e +6\tau_{ph}\over{\tau_e+\tau_{ph}}}.\label{alpha}
\end{equation}

\begin{figure}
\includegraphics[width = 87mm, angle=-0]{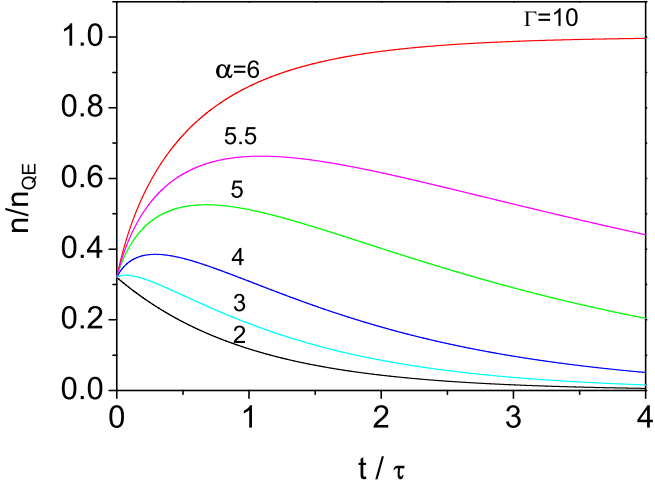}
\caption{Relaxation of the total number of non-equilibrium excitations $n$ for different e-e and e-ph scattering times characterized by  parameter $\alpha$, Eq.(\ref{alpha}), for the  width $\Gamma=10$ of the initial non-equilibrium distribution.}
\end{figure}

There are two bound states of the corresponding steady-state Schr\"odinger-like equation
\begin{equation}
\left [{\partial^2\over{\partial x^2}} +{\alpha\over{\cosh^2(x)}} \right]\psi(x)=-E\psi(x)
\label{soliton3}
\end{equation}
because $\alpha$ is larger than $2$ but smaller than $6$ \cite{landau}. The  excited odd state has the eigenfunction
\begin{equation}
\psi_1(x)\propto {\tanh(x)\over{[\cosh(x)]^{|E_1|^{1/2}}}}
\end{equation}
and the energy
\begin{equation}
E_1=-{1\over{4}}\left[(1+4\alpha)^{1/2}-3\right]^2,
\end{equation}
which determines the asymptotic behavior of all linear response functions. In particular the  number of excitations decays at large $t$ as
\begin{equation}
n(t) \propto \exp \left[-{(1+E_1)t\over{\tau}}\right]. \label{asymp6}
\end{equation}
When both relaxations are involved the time evolution of $n(t)$ calculated using Eq.(\ref{solitonf}) with different initial distributions differs qualitatively   from TTM relaxation as shown in Fig.3 and Fig.4. In fact  electrons cannot attain the thermal
quasi-equilibrium at any $\alpha$ less than $6$ in agreement with the numerical results of Ref.\cite{lag}. Moreover the exact  relaxation rate $\gamma=1+E_1$
depends on the ratio of the electron-electron relaxation time, Eq.(\ref{taue2}), and the electron-phonon  relaxation time,  Eq.(\ref{tauph2}),
\begin{equation}
r={\tau_e\over{\tau_{ph}}}\approx {2\lambda\over {\pi \mu_c^2}} {E_F\over {\hbar \omega_D}}.\label{ratio}
\end{equation}
 Using Eq.(\ref{asymp6}) we find
\begin{equation}
\gamma= c(r){\pi^3\lambda(k_BT)^2\over{2\hbar^2 \omega_D}}\label{ratef},
\end{equation}
where
\begin{equation}
c(r)={3\sqrt{(1+r)(25+9r)}-7r-15\over{2r}}.
\end{equation}
This  coefficient changes from $c(r)=1$ at $r=\infty$ up to $c(r)=8/5$ at $r=0$.

 The TTM relaxation rate $\gamma_{Tlow}$ at low temperatures  is readily obtained with the Eliashberg function $\alpha^2 F(\omega)=\lambda \omega/2\omega_D$ using Eqs. (4,10) of Ref.\cite{allen,ref}.  Linearizing Eq.(10) of Ref. \cite{allen}  with respect to the temperature difference $T_e(t)-T\ll T$ yields
\begin{equation}
\gamma_{Tlow}= {4\pi^3\lambda(k_BT)^2\over{5\hbar^2 \omega_D}}\label{ratef}.
\end{equation}
The ratio of our exact relaxation rate to the TTM rate is
\begin{equation}
{\gamma\over{\gamma_{Tlow}}}= {5\over{8}} c(r)\label{ratef}.
\end{equation}
If  e-e collisions are much faster than e-ph collisions ($r\rightarrow 0$), this ratio is $1$, justifying  the TTM approximation   also at low temperatures in the limit $t \rightarrow \infty, r\rightarrow 0$ . However at low temperatures $r$ is not necessarily small as assumed in TTM even  at small $\lambda \ll \mu_c$ because the Fermi energy in Eq.(\ref{ratio})  is often much larger than the phonon energy. Just the opposite limit $r\rightarrow \infty$ is  feasible at a sizable $\lambda$. In this limit the exact relaxation rate is slower than the low-temperature TTM rate, $\gamma/\gamma_{Tlow}= 5/8$, so that one may underestimate the electron-phonon coupling constant by about two times using TTM. Also an illegitimate fitting of experimental rates measured at temperatures below $\hbar \omega_D/k_B$ with the theoretical high-temperature TTM rate $\gamma_T$ \cite{allen} (see section \ref{sechigh})  may underestimate  $\lambda$ by about $(\hbar\omega_D/\pi k_BT)^{3}$ times in poor metals and much more in clean metals.
\begin{figure}
\includegraphics[width = 87mm, angle=-0]{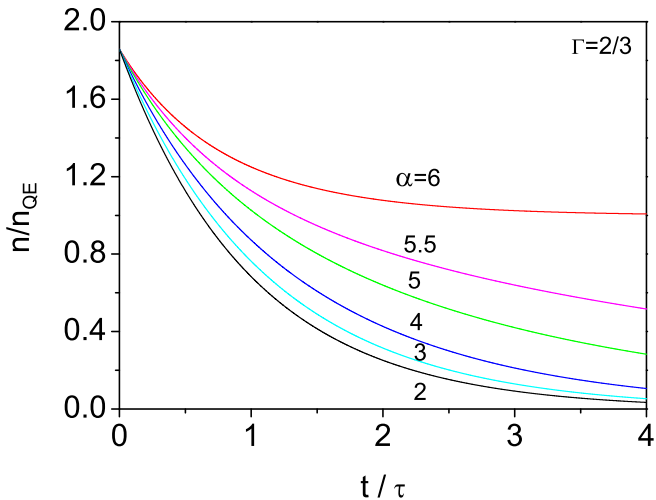}
\caption{Relaxation of the total number of non-equilibrium excitations $n$ for different e-e and e-ph scattering times characterized by  parameter $\alpha$, Eq.(\ref{alpha}), for the  width $\Gamma=2/3$ of the initial non-equilibrium distribution.}
\end{figure}

\section{Conclusions}
In conclusion, using the auxiliary  Fourier transform we have mapped
the linearized Botzmann equation with the electron-electron
collision integral onto a  Schr\"odinger-like  equation with
imaginary time allowing for a simple analytical solution. A similar
mapping is also found for the electron-phonon collision  integral at
low-temperatures both in poor and clean metals. We have analytically
traced the time and energy evolution of the non-equilibrium electron
distribution function on any time scale and found its asymptotic
relaxation rate at $t\rightarrow \infty$.

A low-temperature relaxation rate strongly depends on the
temperature: $\gamma \propto T^2$ and $\gamma \propto T^3$ in poor
and clean metals, respectively. The Pauli exclusion principle slows
down  e-e relaxation, so that  e-e and e-ph collisions are strongly
entangled at low temperatures. We have shown that electron gas
cannot attain a thermal quasi-equilibrium distribution by e-e
collisions, Figs.3,4, in agreement with earlier numerical
integrations of the Boltzmann equation \cite{lag}. The rate of
return to the equilibrium is not governed solely by electron-phonon
processes, but also involves the electron-electron relaxation time,
$\gamma=c(r)\gamma_{Tlow}$, via the coefficient $c(r)$ which depends
on the ratio $r$ of the e-e collision time to the e-ph collision
time. The exact relaxation rate $\gamma$ recovers its
quasi-equilibrium TTM value $\gamma=\gamma_{Tlow}$ only in the
limit of the negligible e-ph coupling, $\lambda \ll \pi \mu_c^2
\hbar \omega_D/2E_F\lesssim  0.01 $ . In poor metals the physically
realistic  ratio $r$  is large at low temperatures  and the exact
relaxation rate is slower than the TTM rate,
$\gamma=5\gamma_{Tlow}/8$.

At high temperatures, $T \gg \hbar \omega_D/k_B$, we have reduced
the e-ph collision integral to the differential Landau-Fokker-Plank
form. Using this form we have shown that the deviation of the
electron distribution from quasi-equilibrium population does not
have much influence on the energy relaxation, so that TTM
\cite{kag,allen} is a reliable approximation at high temperatures.

Our  theory  opens up a perspective of determinations of both
important microscopic parameters $\lambda$ and $\mu_c$ using
single-parameter $(\alpha)$ fitting of response functions in
pump-probe experiments at low temperatures, Figs. 3,4. It also
allows for an  analytical approach to the integral Boltzmann
equation for the case of a steady-state source of excitations as
 in a current-carrying state. In the latter case relaxation
times could be different because the  current carrying state does
not have a distribution function that depends only on energy, as
assumed here. The theory could be further extended beyond the
assumption that phonons remain in equilibrium by including a
linearised Boltzmann equation for the non-equilibrium phonon
distribution function.

We are grateful to  Alexander Veselov and Rajmund Krivec  for
illuminating  discussions of  Calogero-Moser-Sutherland  equations
and to Dragan Mihailovic for sharing with us his insight into
non-equilibrium phenomena in metals and high-temperature
superconductors.   The work was supported by the Slovenian Research
Agency (ARRS) (grant no. 430-66/2007-17) and by EPSRC (UK) (grant
no. EP/D035589/1).


\begin{thebibliography}{200}
\bibitem{ref} We have noticed that a multiplier $\pi^2 N_c N(E_F)/3$ is missing in the second term of Eq.(4) in Ref. \cite{allen}.
\end{thebibliography}
\end{document}